\documentclass[twocolumn,prl,twoside,a4paper]{revtex4}
\usepackage{graphicx}
\usepackage{amssymb}

\begin{document}

\title{Recovering correlations in heterodyne spectra for high-precision quantum displacement sensing}

\author{ T.~S.~Monteiro} 
\email{t.monteiro@ucl.ac.uk}
\author{J.E. Lang}
\address{Department of Physics and Astronomy, University College London, Gower Street, London WC1E 6BT, United Kingdom}
\email{t.monteiro@ucl.ac.uk}

\begin{abstract}
Homodyne and heterodyne detection represent  ``twin-pillars'' of quantum 
displacement sensing using optical cavities, having permitted major breakthroughs including 
detection of gravitational waves and of the motion of quantum ground-state cooled mechanical oscillators. 
Both can suffer disadvantages as diagnostics in quantum optomechanics, either through symmetrisation (homodyne), or loss of correlations (heterodyne). We show that, for  modest
heterodyne beat frequencies ($\Omega \sim \omega_M/10 \gg \Gamma$),  judicious construction of the autocorrelation of the measured current can either recover (i)  a spectrum with strong sidebands 
 but without an imprecision noise floor (ii)  a spectrum which is a hybrid,
combining both homodyne and heterodyne sideband features. We  simulate an experimental realisation with stochastic numerics and find excellent agreement with analytical quantum noise spectra.  We term such retrospective recovery of lost heterodyne correlations ``r-heterodyning'':  as the method simply involves post-processing of a normal heterodyne time signal, there is no additional experimental constraint other than on the magnitude of $\Omega$.
\end{abstract}

\maketitle 

The extraordinary sensitivity of cavity-mediated detection was demonstrated by the recent LIGO detection
of gravitational waves \cite{LIGO} which sensed displacements even down to  order $10^{-18}$ m. Earlier versions of LIGO employed a radio-frequency (RF) heterodyne detection system, but this was later replaced by a homodyne scheme
\cite{LIGODC}. The related field of quantum cavity optomechanics has also exposed a rich seam of interesting phenomena
arising from the coupling between the mode of a cavity and a small mechanical oscillator
\cite{Bowenbook,AKMreview}. Several groups have successfully cooled a mechanical oscillator 
\cite{Teufel2011,Chan2011,Kipp2012}  down to  mean phonon occupancy $\overline{n} \sim 1$ or under,
 close to its quantum ground state. Read-out of the temperature was achieved by detection of motional sidebands in the cavity output by homodyne or heterodyne methods.

An important herald of the quantum regime of optomechanics is the appearance of asymmetric displacement sidebands; these have been studied experimentally  \cite{SideAsymm2012,SideAsymm2012a,Weinstein}.  Albeit indirectly \cite{SideAsymm2012a,Weinstein}, the observations mirror an underlying asymmetry in the  motional spectrum: an oscillator in its ground state $\overline{n}=0$, can absorb a phonon and down-convert the photon frequency (Stokes process); but it can no longer emit any energy and up-convert a photon (anti-Stokes process). Heterodyne detection is the most effective method to detect sideband asymmetry and has also been used to establish  cooling limited by only quantum backaction \cite{Peterson}. Homodyne detection on the other hand allows detection of ponderomotive squeezing, whereby narrowband cavity output falls below the technical imprecision noise  is also of much current interest \cite{Safavi2013,Purdy2013,Pontin} since the noise floor arising from optical shot noise limits detection of small displacements. 

In the present work we introduce and investigate a post-processing method which yields a hybrid of the homodyne and heterodyne features; and moreover can improve the signal to noise ratio  hence potentially offers a ``best of both worlds'' scenario.

\begin{figure*}[ht]
\begin{center}
{\includegraphics[width=6.3in]{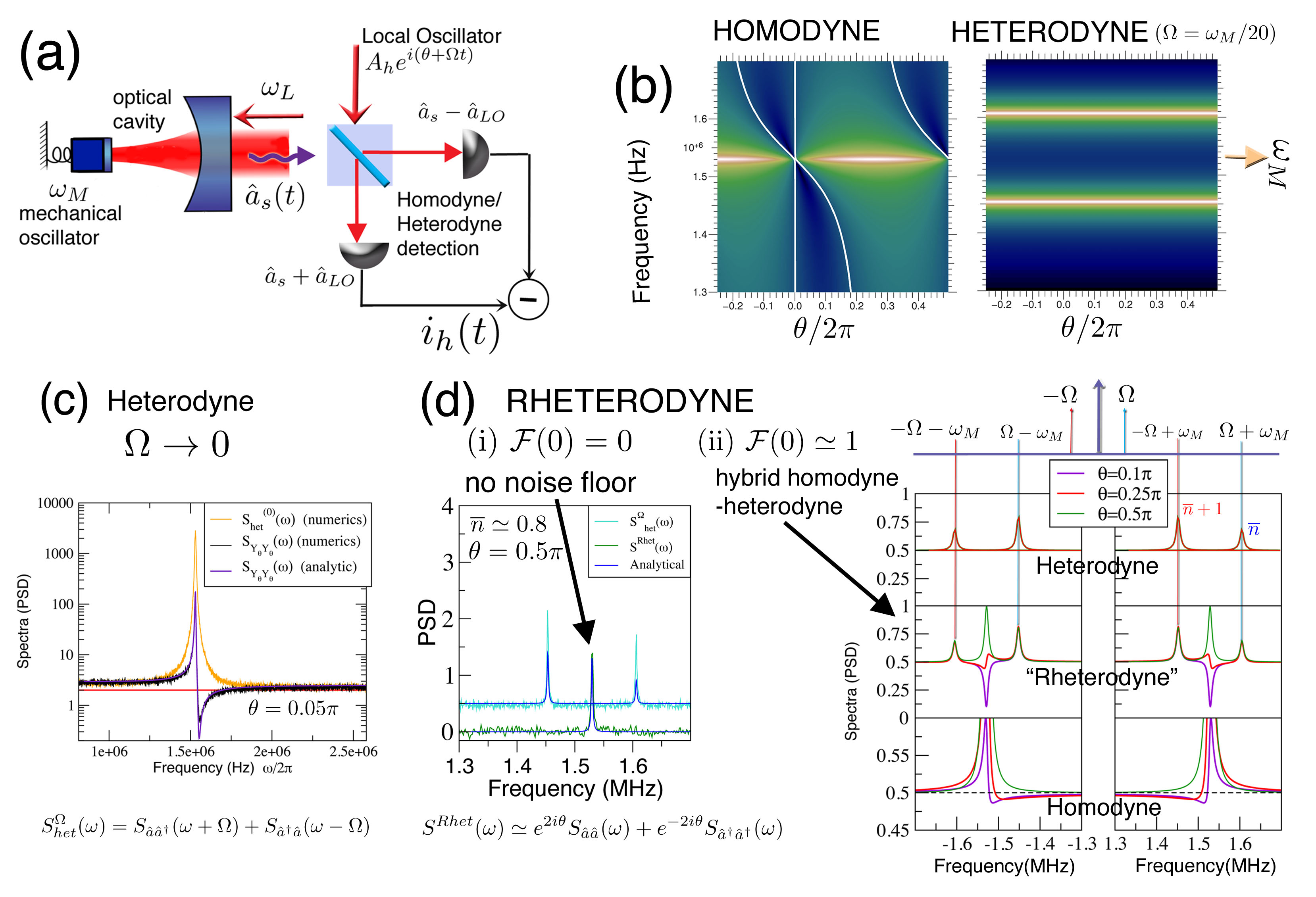}}
\end{center}
\caption{ {\bf (a)} The output field from an optical cavity provides exquisitely sensitive detection of quantum-scale displacements. Balanced detection  of the beat between  signal and a local oscillator of frequency $\Omega$ and phase $\theta$ is termed heterodyne for $\Omega \neq 0$ and homodyne for $\Omega=0$. The signal is modulated by the dynamics of interest yielding sidebands  near characteristic frequencies $\pm \omega_M$. {\bf (b)} Illustrates key features of the PSD of of the detected current $i_h(t)$ (on a log scale). For homodyning, a single optical quadrature 
$\hat{Y}_\theta$ is measured and ponderomotive squeezing can (over a narrow frequency range) yield PSD  values below the imprecision noise floor (the region between white lines). In  contrast, the PSD of the heterodyne spectra  $S^{\Omega}_{het}(\omega)$ shows neither squeezing nor $\theta$ dependence but now has two sidebands at 
$\omega \simeq \pm (\Omega\pm \omega_M)$. {\bf (c)} We compare  stochastic numerics
  with analytical noise spectra. The transition from heterodyne to homodyne as $\Omega \to 0$ is too abrupt to be  typically observable: as $\Omega \to 0$, the two heterodyne peaks  merge before any evidence of coherences or $\theta$ dependence become apparent. $S^{0}_{het}(\omega)\equiv S^{\Omega \lesssim  \Gamma}_{het}(\omega)$ 
corresponds to a single peak, with no squeezing.  
{\bf (d)}  Illustrates rheterodyning:  application of a filter function  $F(t)=FT^{-1}[\mathcal{F}(\omega)]$ when constructing the PSD of   $i_h(t)$ yields instead $S^{Rhet}(\omega) = \mathcal{F}(0)S^{\Omega}_{het}(\omega)+ \mathcal{F}(2\Omega)[S^{hom}_{Y_\theta Y_\theta}(\omega)-S^{0}_{het}(\omega)] $. 
(i) A key result is that choosing a filter with $\mathcal{F}(0)=0$ yields strong sidebands but no significant imprecision noise, in contrast with homodyne and heterodyne PSDs, which in the quantum regime exhibit a flat, half-quantum, shot noise floor. (ii)The effect of a filter with $\mathcal{F}(0) \simeq \mathcal{F}(2\Omega))\simeq 1$ is shown on the right and compared with usual heterodyne/homodyne PSDs. Here the two peaks of the usual heterodyne PSD are present
and the PSD simultaneously shows sideband asymmetry (about $\pm \Omega$) as well as squeezing. 
 Since the rheterodyne PSD  subtracts the heterodyne and homodyne PSDs it  takes negative
values near $\theta\approx 0$.  We emphasize that the rheterodyne and heterodyne spectra above use the {\em same} time trace $i_h(t)$: rheterodyning is  a post-processing method.
Parameters similar to experiments in \cite{Purdy2013} were used in both {\bf (b)} and {\bf (c)}, with a minor adjustment in {\bf (d)}
to obtain steady-state phonon occupancies of $\overline{n}\simeq 1$,
see Appendix for details of parameters).}
\label{Fig1}
\end{figure*}

As illustrated in Fig.\ref{Fig1}(a), in balanced homodyne or heterodyne detection the output field of a cavity is amplified by beating with a reference local oscillator $A e^{i(\Omega t+\theta)}$ of amplitude $A$ much larger than cavity output amplitudes,  and is subsequently detected as a current 
which, normalised to $A$, takes the form: 
\begin{equation}
{\hat i}_h(t)= {\hat{a}}^\dagger e^{i(\Omega t+\theta)}+ {\hat{a}}e^{-i(\Omega t+\theta)}
\end{equation}
 and its experimental investigation usually proceeds via the power spectral density (PSD):
\begin{eqnarray}
\langle |{\hat i}_h(\omega)|^2\rangle=\frac{1}{T}\int_0^T dt'\int_0^T dt \langle {\hat i}^\dagger(t) {\hat i}(t')\rangle e^{i\omega(t-t')}.
\label{FTauto}
\end{eqnarray}
The Wiener-Kinchin theorem simplifies calculation of the PSD of an arbitrary observable $Q$, which is stationary and in its steady state, to a Fourier transform of  its  autocorrelation function:
$\langle |{\hat Q}(\omega)|^2\rangle \equiv S_{{\hat Q}^\dagger {\hat Q}}(\omega)
= \int_0^T ds \langle {\hat Q}^\dagger(s) {\hat Q}(0)\rangle e^{i\omega s}$.
  For the optomechanical current  ${\hat i}^\dagger_h={\hat i}_h$. The
 actual measured current $ \equiv i_h(t_k)$, for $k=1,2,3..N_T$, is obtained as a
 discrete time series.

In optomechanics experiments,  the interesting dynamical features appear in frequency space as peaks (sidebands)  near  $\omega \simeq \pm \omega_M$  where $\omega_M$ is the mechanical frequency. The sidebands have width of order $\Gamma=\gamma_M+\Gamma_{opt}$, where $\gamma_M$ is the intrinsic  mechanical damping  and $\Gamma_{opt}$ is the optical 
damping.

\begin{figure*}[ht]
\begin{center}
{\includegraphics[width=6.4in]{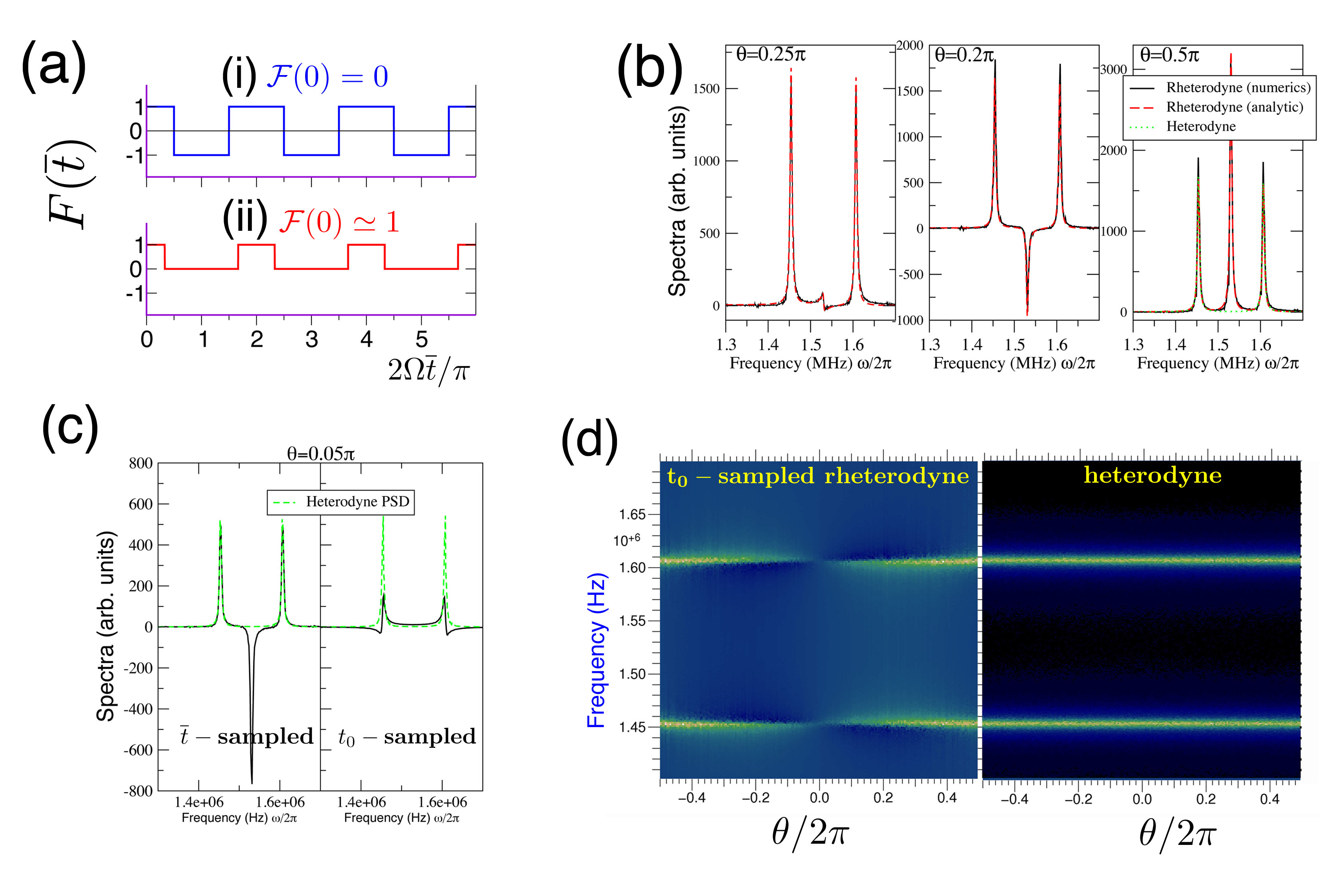}}
\end{center}
\caption{
  {\bf (a)} Form of the filter functions $F(\overline{t})$ which in frequency space correspond
to the desired  (i) $\mathcal{F}(0)=0$ where we toggle $F(\overline{t})$ between $\pm1$ values (ii) $\mathcal{F}(0)\simeq \mathcal{F}(2\Omega)$ where $F(\overline{t})=1$ for 
$2\Omega \overline{t}=2\pi N \pm \pi/3$.  The case (ii) filter shifted by $\pi/2$ reproduces closely the full heterodyne result, so can be used to identify the phase of the Local Oscillator at a given time, if unknown in an experimental heterodyne trace. 
 {\bf (b)} For the type (ii) filter, shows comparisons between
 analytical calculation of $S^{Rhet}(\omega)= S^{\Omega}_{het}(\omega)+ [S^{hom}_{Y_\theta Y_\theta}(\omega)-S^{0}_{het}(\omega)] $ 
from the quantum noise spectra and semiclassical numerically calculated Fourier transforms of ${\overline t}$-sampled autocorrelation functions. The steady state phonon number $\overline{n}=90$ and agreement 
 is excellent. {\bf (c)} Compares the effect of three types of post-processing on the {\em same}
numerically calculated time trace for the heterodyne current: (1)  a standard Fourier transform (green line) yielding the normal heterodyne spectrum;
(2) a ${\overline t}$-sampled rheterodyning (left panel; correlations are isolated) and (3) $t_0$-sampled rheterodyning (right panel;
correlations interfere with the main heterodyne peaks).
 {\bf (d)} Map compares $t_0$-sampled rheterodyne spectra with ordinary heterodyne spectra.
 For $t_0$-sampled rheterodyning the recovered correlations interfere with the
main heterodyne peaks so the latter become $\theta$-dependent.}
\label{Fig2}
\end{figure*}

Homodyne detection corresponds to $\Omega=0$ so leads to detection of a single quadrature 
$\hat{Y}_\theta={\hat{a}}^\dagger e^{i\theta}+ {\hat{a}}e^{-i\theta}$  and the corresponding PSD
 of the current, $ S^{hom}_{Y_\theta Y_\theta}(\omega)$,
 may be written:
\begin{eqnarray}
S^{hom}_{Y_\theta Y_\theta}(\omega) &\equiv& S_{\hat{a}^\dagger \hat{a}}(\omega)+S_{\hat{a} \hat{a}^\dagger}(\omega)\nonumber\\
            & +& e^{-2i\theta} S_{\hat{a} \hat{a}}(\omega)+S_{\hat{a}^\dagger \hat{a}^\dagger}(\omega)e^{2i\theta} 
\label{homo}
\end{eqnarray}
where e.g. $S_{{\hat a} {\hat a}}(\omega) \equiv \int_0^T ds \langle {\hat a}(s) {\hat a}(0)\rangle e^{i\omega s}$. The last two terms in Eq.\ref{homo} represent coherent contributions which make the homodyne PSD strongly depend on the measured quadrature.
As shown in Fig.\ref{Fig1}(b) the PSD varies sharply with $\theta$ and its value drops below the technical imprecision
noise floor (region  between the white lines in the figure), indicating squeezing of the light field due to the mechanical motion (ponderomotive squeezing). The measured current is the single-sided symmetrised version 
$\overline{S}^{hom}_{Y_\theta Y_\theta}(\omega)=\frac{1}{2}[S^{hom}_{Y_\theta Y_\theta}(\omega)+S^{hom}_{Y_\theta Y_\theta}(-\omega)]$.
In contrast, for $\Omega > 0$, heterodyne detection yields the PSD:
\begin{equation}
S^{\Omega}_{het}(\omega)= S_{\hat{a}\hat{a}^\dagger}(\omega+\Omega)
              +S_{\hat{a}^\dagger\hat{a}}(\omega-\Omega)
\label{heterodynePSD}
\end{equation}
for experimental time-traces where $T \gg 2\pi/\Omega$. 
The rotating quadratures $\hat{Y}_{het}=\hat{a} e^{-(i\theta+\Omega t) }+\hat{a}^\dagger e^{i(\theta+\Omega t)}$ 
lead to averaging out in time and loss of the $S_{\hat{a}\hat{a}}$ 
and $S_{\hat{a}^\dagger\hat{a}^\dagger}$ components, as may be seen by comparison 
with Eq.\ref{homo} and there is no longer any dependence on $\theta$.
In addition, the sidebands are shifted:
 if the cavity output is modulated by dynamics of characteristic frequencies $\pm\omega_M$,
the heterodyne current exhibits sidebands peaked near $\omega=\Omega \pm \omega_M$ and $\omega=-\Omega \pm \omega_M$. To resolve them, we need $T\gg \Gamma^{-1} \gg 2\pi/\Omega$.

Fig 1(b) shows also that while the  PSDs of the homodyne current exhibit a strong dependence on $\theta$, in contrast, the heterodyne spectra are completely insensitive to $\theta$. However, the  PSDs of the heterodyne current yield  two, rather than one, sidebands near $\omega \simeq +\omega_M$. Although 
$S^{\Omega}_{het}(\omega)=S^{\Omega}_{het}(-\omega)$  
the sidebands are not necessarily symmetric about
$\Omega$: Stokes/anti-Stokes sideband asymmetry, is thus conveniently observable providing a clear quantum signature \cite{Peterson,KippNun2016}. 

Comparison between Eq.\ref{heterodynePSD} and Eq.\ref{homo} shows that the first equation does not reduce to the latter as $\Omega \to 0$. This reflects the fact that the transition is abrupt: since  $T \gg 2\pi/\Omega$, the additional correlations are fully averaged out.
Optomechanics experiments do not easily probe the heterodyne-homodyne transition. Since  $T   \gg \Gamma^{-1} \gg  2\pi/\Omega$
 the heterodyne sidebands of Fig.\ref{Fig1}(b) become unresolvable as $\Omega \to 0$, 
 before any correlations of the true $\Omega=0$ limit are manifest in the spectrum.
Hence one may consider a $\Omega \to 0$ {\em heterodyne} spectrum:
\begin{equation}
S^{0}_{het}(\omega)= S_{\hat{a}\hat{a}^\dagger}(\omega)
              +S_{\hat{a}^\dagger\hat{a}}(\omega)
\label{heterodynePSD}
\end{equation}
which would be measured after the merging of sidebands as $\Omega \to \Gamma$, but devoid of additional correlations; there is simply a single peak which is invariant with $\theta$. 
But for any $\Omega$, the loss of correlations here arises externally, purely in the heterodyne time-averaging stage. The underlying intra-cavity fluctuations are the same, and 
 in typical optomechanical experiments, are stationary (time-translationally invariant).

Hence here we show  that for arbitrarily long $T$ but modest
$ \Omega \sim \omega_M/10-\omega_M/100 \gg \Gamma$ there is a robust and completely straightforward 
procedure to recover the lost correlations. Our approach is to recast Eq.\ref{FTauto} in the form:
\begin{eqnarray}
\langle |{\hat i}_h(\omega)|^2\rangle=\frac{1}{T}\int_0^T  F(\overline{t})  d\overline{t}
 \int_0^T\langle i(\overline{t}-s/2)i(\overline{t}+s/2) \rangle e^{i\omega s}ds
\label{Rhetav}
\end{eqnarray}
where we used $s=t-t'$ and $\overline{t}=(t+t')/2$. Above, $F(\overline{t})$ is a filter function that we choose for convenience. We define its Fourier transform $\textrm{FT}[F(t)]\equiv \mathcal{F}(\omega)$
and restrict ourselves here to cases for which $\mathcal{F}(\omega)=\mathcal{F}(-\omega)$. 
It is simple to show (see \cite{Suppinfo}) that we can approximate 
$\langle |{\hat i}_h(\omega)|^2\rangle\equiv S^{Rhet}(\omega)$
by:
\begin{equation}
S^{Rhet}(\omega)\simeq \mathcal{F}(0)S^{\Omega}_{het}(\omega)+ \mathcal{F}(2\Omega)[S^{hom}_{Y_\theta Y_\theta}(\omega)-S^{0}_{het}(\omega)].
\label{rheterodyne}
\end{equation}
We investigate here two cases: (i) the behaviour with a filter function for which $\mathcal{F}(0)=0$
 (ii) the behaviour for the case $\mathcal{F}(0)\simeq \mathcal{F}(2\Omega)$.
As shown in Fig.\ref{Fig1}(d), for case (i) the usual half-quantum imprecision due to quantum shot
noise is absent from the spectrum, since $[S^{hom}_{Y_\theta Y_\theta}(\omega)-S^{0}_{het}(\omega)]
\equiv e^{-2i\theta} S_{\hat{a}\hat{a}}(\omega) +e^{2i\theta} S_{\hat{a}^\dagger\hat{a}^\dagger}(\omega)$;  the terms  
$S_{\hat{a}\hat{a}}(\omega)$ and $S_{\hat{a}^\dagger\hat{a}^\dagger}(\omega)$ do not carry this noise floor. However, the sidebands of the filtered spectrum
can be stronger (e.g. for $\theta \approx \pi/2$) than heterodyne peaks obtained the usual way. These features are of clear interest for sensing applications. 
Case (ii) includes both heterodyne and homodyne features; as the $S^{0}_{het}(\omega)$
is a simple $\theta$ invariant peak,  it can be estimated
or measured, and hence the  homodyne spectrum $S^{hom}_{Y_\theta Y_\theta}(\omega)$ 
 can also be recovered to a good approximation.  As seen in Fig.\ref{Fig1}(d) one may 
obtain a spectrum (middle panel) that simultaneously exhibits quantum sideband asymmetry as well as quantum squeezing: the recovered $\theta$ dependence is evident from the plots.

The time-domain form of the filter functions $F(\overline{t})$ which correspond to cases (i) and (ii)
in the frequency domain,
are illustrated in Fig.\ref{Fig2}(a) along with the construction of Eq.\ref{Rhetav}, and are defined as follows:\\
 (i) $ F(\overline{t})=+1$ for  $2\Omega \overline{t} =2\pi N +\phi$ with $\phi\in[-\pi/2:\pi/2]$ and $F(\overline{t})=-1$ elsewhere for $N=0,1,2...$   \\
(ii) $F(\overline{t})=+1$ for  $2\Omega \overline{t}= 2\pi N+\phi$ with $\phi\in[-\pi/3:\pi/3] $ but $F(\overline{t})=0$ elsewhere.

We test Eq.\ref{Rhetav} against explicit stochastic simulation of
a heterodyne current time series $i_h(t_k)$, for $k=1,2,3...N_T$ akin to what would be measured experimentally. We can then compare  with analytical $S^{Rhet}(\omega)$ constructed 
from standard quantum noise spectra solutions of Eqs.\ref{homo},\ref{heterodynePSD} and 
Eq.\ref{rheterodyne}. The stochastic Langevin numerics were previously developed to simulate
levitated optomechanics experiments \cite{PRL2016,Aranas2016}. Here, they 
are refined to accurately preserve correlations between input noise to the cavity and the cavity filtered noises  making the computations more laborious, ($N_T \sim 10^6$ with typically 100 stochastic realisations). We include two optical modes (for cooling, one for detection)  and obtain the analytical quantum noise spectra for the two-mode
problem as in \cite{Monteiro2013}.

Excellent quantitative agreement was obtained, especially for input optical noise ${\hat a}_{in}$ levels  slightly
higher  than shot noise (e.g. taking $n_p = 2$ in $\langle {\hat a}_{in}^\dagger(t') {\hat a}_{in}(t) \rangle =  \bar n_p \ \delta(t-t')$
rather than $n_p=0$) in both the semiclassical stochastic numerics and the quantum analytical solutions.
Fig.\ref{Fig2}(b) compares case (ii) numerics and analytical calculation of  
$S^{Rhet}(\omega)\simeq S^{\Omega}_{het}(\omega)+ [S^{hom}_{Y_\theta Y_\theta}(\omega)-S^{0}_{het}(\omega)] $. The signal to noise ratio is of of course worse by a corresponding factor of 2-3 or so since the case (ii) numerics use only $1/2-1/3$rd of the values of $\overline{t}$.

In contrast, case (i), as shown in Fig.\ref{Fig1}(d) has far higher signal to noise ratio: not only is 
all the data used, but the background shot noise floor is absent.
While practical implementation would ideally involve storing the LO phase with time in each experiment, this is not essential : the method involves post-processing of a standard heterodyne output and 1D Fourier transforms are computationally simple; for case (ii) one searches for the filter initial  phase which best reproduces the full heterodyne traces; then a simple $\pi/2$  shift of $F(\overline{t})$  represents the rheterodyne case.

Interestingly,  the rheterodyned PSD of  $i_h(t)$ constructed using the usual `initial-time' form of the autocorrelation  (taking the substitutions $t=t_0+s$ and $s=t-t'$)  gives completely different behaviour. Writing:
 \begin{eqnarray}
\langle |{\hat i}_h(\omega)|^2\rangle=\frac{1}{T} \int_0^T F(t_0) dt_0 \int_0^T ds \langle i(t_0)i(t_0+s) \rangle e^{i\omega s}
\label{Rhet0}
\end{eqnarray}
 we find the lost correlations are then shifted  by $\pm \Omega$ and coincide with the 
main heterodyne peaks. In contrast, for standard heterodyne, the change from $\overline{t}$ 
to the (more usual) $t_0$ is a trivial and the PSDs completely equivalent.
 Fig.\ref{Fig2}(c) and (d) illustrates the effect, on exactly the same measured current of (1) a standard heterodyne (Fourier transform
of the full autocorrelation function (2) $\overline{t}$-sampled rheterodyne (3) $t_0$-sampled rheterodyne as in Eq.\ref{Rhet0}. While for the former the correlations are isolated from the 
usual heterodyne sidebands, for the latter they coincide and interfere with the main peaks. 
However, the $t_0$-sampled PSD is harder to understand in terms of the underlying heterodyne and homodyne constituents: comparison with corresponding analytical forms are less quantitatively accurate 
relative to the isolated correlations of the $\overline{t}$-sampled case so details of this interference are not fully understood yet and need further investigation.

{\em Conclusions} In the present work we propose a procedure which, despite its simplicity, qualitatively augments  the information and sensitivity obtainable in optomechanical displacement sensing. It also offers an alternative probe of dynamics which modulated the  cavity output. We argue that this extremely simple post-processing method
is straightforward to realise experimentally. Implications for quantum sensing will be explored 
further in future work.  
Acknowledgements: the authors acknowledge useful discussions with Erika Aranas.
The work was supported by EPSRC grant EP/N031105.

\section*{Appendix}
\subsection{Parameters for calculations}
In Fig.\ref{Fig1} we use the parameters close to those in the experiments in  \cite{Purdy2013}. Hence 
$\omega_M/2\pi=1.53$ MHz, $\kappa=1.7$ MHz and $\gamma_M=0.22$ Hz while $T_B=4.6$ K.
This is a two-mode scheme, with a probe beam and a
damping beam. The optomechanical coupling for the damping beam is taken to be $G_d/2\pi=3.5 \times 10^4$ Hz 
while its detuning $\Delta_p=-1.53$ MHz so the damping beam provides an optical damping of order $2700$ Hz. 
For the probe beam, $\Delta=-21$KHz while $G/2\pi=3.5 \times 10^5$ Hz.
In Fig.\ref{Fig1}(d) the probe strength was reduced to  $G/2\pi=3.5 \times 10^4$ Hz and $\gamma_M=0.22d-1$ Hz to allow cooling to $\overline{n}=0.8$.
The same values were used in Fig.\ref{Fig2}. However to make comparisons with semiclassical Langevin numerics easier,
the photon bath number was taken to be $n_p=2$ instead of $n_p=0$ for shot noise.
\subsection{PSD of the current with filter function}
If we calculate the PSD of the measured current from the experimental time-series with a filter function 
we have:
\begin{eqnarray}
\langle |{\hat i}_h(\omega)|^2\rangle=
\frac{1}{T}\int_0^T  F(\overline{t})  d\overline{t}&
 \int_0^T\langle i(\overline{t}-s/2)i(\overline{t}+s/2) \rangle e^{i\omega s}ds\nonumber\\
\end{eqnarray}
In terms of the cavity fields this becomes:
\begin{eqnarray}
&T \ & \langle |{\hat i}_h(\omega)|^2\rangle=\nonumber\\
&\int_0^T & F(\overline{t})  d\overline{t}
 \int_0^T\langle a(\overline{t}-s/2)a^\dagger(\overline{t}+s/2)+ \rangle e^{i(\omega-\Omega) s}ds\nonumber\\
&\int_0^T & F(\overline{t})  d\overline{t}
 \int_0^T\langle a^\dagger(\overline{t}-s/2)a(\overline{t}+s/2)+ \rangle e^{i(\omega+\Omega) s}ds\nonumber\\
&\int_0^T  & F(\overline{t}) e^{2i(\Omega \overline{t}+\theta)} d\overline{t}
 \int_0^T\langle a(\overline{t}-s/2)a(\overline{t}+s/2)+ \rangle e^{i\omega s}ds\nonumber\\
&\int_0^T&  F(\overline{t}) e^{-2i(\Omega \overline{t}+\theta)} d\overline{t}
 \int_0^T\langle a^\dagger(\overline{t}-s/2)a^\dagger(\overline{t}+s/2)+ \rangle e^{i\omega s}ds\nonumber\\
\label{IntSaa}
\end{eqnarray}
Clearly  $F(\overline{t})=1$ is the usual heterodyne spectrum. We then simply note that all the integrals over $ds$ involve time translationary-invariant integrals hence we can use the usual Wiener-Kinchin approximation and replace e.g. the first integral by 
$S_{a a^\dagger}(\omega)= \int_0^T\langle a(s/2)a^\dagger(s/2) \rangle e^{i\omega s}ds$ etc.
The integrals are then separable; the  left-hand side integrals are replaced by the appropriate  FTs of $F(t)$. Hence we obtain:
\begin{eqnarray}
 \langle |{\hat i}_h(\omega)|^2\rangle \simeq
 \frac{1}{T} \mathcal{F}(0)[S_{\hat{a}^\dagger \hat{a}}(\omega-\Omega)+S_{\hat{a} \hat{a}^\dagger}(\omega+\Omega)] \nonumber\\
               +  \frac{1}{T} \mathcal{F}(2\Omega)[e^{-2i\theta} S_{\hat{a} \hat{a}}(\omega)+S_{\hat{a}^\dagger \hat{a}^\dagger}(\omega)e^{2i\theta} ]
\label{retro}
\end{eqnarray}
The `rheterodyne' expression in the main text. We then seek appropriate filter functions. To obtain the case (ii) hybrid homodyne-heterodyne spectra
of Fig.\ref{Fig2} we want $\mathcal{F}(0)\simeq \mathcal{F}(2\Omega)\simeq 1$. 

An additional normalisation may be required since for example, 
the case (ii) we use only $1/3$rd of the available values of $\overline{t}$ so would rescale by a factor $3$ to obtain sidebands of equivalent height to the ordinary heterodyne PSD (we also  wish to correct for the sinc-function form of $\mathcal{F}(\omega)$ hence $\mathcal{F}(2\Omega)\lesssim 1$). The result is a higher noise floor. This is a negligible effect in the thermal regime of Fig.\ref{Fig2}(b) but is  important in the quantum sensing regime.

For quantum sensing, case (i) is most appropriate since then we have:
 \begin{eqnarray}
 \langle |{\hat i}_h(\omega)|^2\rangle \simeq  \frac{1}{T} 
 \mathcal{F}(2\Omega)[e^{-2i\theta} S_{\hat{a} \hat{a}}(\omega)+S_{\hat{a}^\dagger \hat{a}^\dagger}(\omega)e^{2i\theta} ]\nonumber\\
 \simeq S_{\hat{a} \hat{a}}(\omega)+S_{\hat{a}^\dagger \hat{a}^\dagger}(\omega)
\end{eqnarray} 
for the case $\theta=\pi/2$. These represent spectra with stronger sidebands than the heterodyne case; they are still less strong than the homodyne case which is (for the phase quadrature) 4 times stronger than heterodyne. However the latter have an imprecision noise floor even for the case of a shot-noise limited laser
whereas the $S_{\hat{a} \hat{a}}(\omega)+S_{\hat{a}^\dagger \hat{a}^\dagger}(\omega)$ spectra are real, but have no imprecision floor so offer promising  possibilities for enhancing displacement sensing in the quantum regime.

For the case of initial time, $t_0$-sampled spectra, we begin from Eq.\ref{FTauto} in the main text:
\begin{eqnarray}
\langle |{\hat i}_h(\omega)|^2\rangle=\frac{1}{T}\int_0^T dt'\int_0^T dt \langle {\hat i}^\dagger(t) {\hat i}(t')\rangle e^{i\omega(t-t')}.
\end{eqnarray}
and use the substitution $t=t'+t_0$ as well as $t-t'=s$.  In that case we obtain an expression very similar to Eqs.\ref{IntSaa} and \ref{retro} except that we have:
\begin{eqnarray}
  S_{\hat{a} \hat{a}}(\omega)&\to& S_{\hat{a} \hat{a}}(\omega-\Omega)\nonumber\\
  S_{\hat{a}^\dagger \hat{a}^\dagger}(\omega)&\to &S_{\hat{a}^\dagger \hat{a}^\dagger}(\omega+\Omega)
\end{eqnarray}
so that the restored correlations coincide in frequency with the main heterodyne peaks.

 \end{document}